\documentclass[%
 preprint,
nofootinbib,
 amsmath,amssymb,
 aps,
showkeys,
prd,
]{revtex4-2}

\usepackage[hidelinks,colorlinks,citecolor=blue,urlcolor=blue,linkcolor=blue]{hyperref}

\usepackage{natbib}
\usepackage{CJK}
\usepackage{graphicx}
\usepackage{dcolumn}
\usepackage{bm}

\usepackage{orcidlink}

\begin{document}


\title{FROLOV BLACK HOLE SURROUNDED BY A CLOUD OF STRINGS}

\author{F. F. NASCIMENTO \orcidlink{0009-0009-7943-5368}}
\email{fran.nice.fisica@gmail.com}
\affiliation{Departamento de F\'isica, Universidade Federal da Paraíba, Caixa Postal 5008 Jo\~ao Pessoa, PB, CEP 58051-970, Brazil}

\author{J. C. ROCHA \orcidlink{0000-0003-4515-9245}}
\email{julio.rocha@servidor.uepb.edu.br}
\affiliation{Departamento de F\'isica, Universidade Estadual da Paraíba Campina Grande, PB, CEP 58429-500, Brazil }
\author{V. B. BEZERRA \orcidlink{0000-0001-7893-0265}}
\email{valdir@fisica.ufpb.br}
\affiliation{Departamento de F\'isica, Universidade Federal da Paraíba, Caixa Postal 5008 Jo\~ao Pessoa, PB, CEP 58051-970, Brazil}
\author{J. M. TOLEDO \orcidlink{0000-0001-9284-0549}}
\email{jefferson.m.toledo@gmail.com}
\affiliation{Departamento de F\'isica, Universidade Federal da Paraíba, Caixa Postal 5008 Jo\~ao Pessoa, PB, CEP 58051-970, Brazil}
%


\date{\today}

\begin{abstract}
We obtain the metric which describes the spacetime corresponding to the Frolov black hole in the presence of a cloud of strings and discuss how  this cloud affects the regularity of the solution and the energy conditions. In addition, we analyze geodesics, effective potential, and several thermodynamic aspects. Finally, we compare our results with the corresponding findings in the literature for the original Frolov black hole, that is, in the absence of a cloud of strings
\end{abstract}

\keywords{Regular black holes;Frolov black hole;
Cloud of strings}
\maketitle

\section{Introduction}
\label{sec:Intro}
Black holes are regions of space where gravity is so intense that even light cannot escape its pull. It forms when extremely dense masses collapse under their own weight, creating an event horizon that marks the boundary between what can be seen and what remains hidden.

Studying black holes helps us understand the evolution of galaxies, since many are located at their centers, influencing the dynamics of stars. Through indirect observations such as radiation from accretion disks and gravitational waves generated by black hole mergers, we can expand our knowledge of how the universe works on cosmic scales. 

The first and simplest solution for a black hole was proposed by Schwarzschild in 1916 \cite{schwarzschild1916uber} through the solution of Einstein's equations for a spherically symmetric static body. A generalization of the solution to include an electric charge was made independently by Reissner (1916) \cite{reissner1916eigengravitation} and Nordström (1918) \cite{nordstrom1918een}.
 In 1963, Kerr derived a stationary metric that describes a rotating black hole \cite{kerr1963gravitational}. Two years later, the solution for a stationary charged black hole, known as the Kerr-Newmann solution \cite{newman1965metric}, was obtained.
 
 All these black holes have singularities at their centers and are therefore known as singular black holes. The presence of singularities implies that the laws of physics are no longer well defined at that point, which may represent a limitation of the theory since at that point there is an incompleteness of geodesic motion and physical quantities also diverge.

 In order to solve this problem related to singularity, regular (non-singular) black holes arise, which are solutions that do not have singularity at $r=0$, consequently, curvature and physical quantities remain finite throughout space-time.

 The first model for a regular black hole was proposed by  Bardeen \cite{bardeen1968non} in 1968. The original solution describes a core of finite density that avoids singularity at the origin. Currently, there are several metrics for static regular black holes \cite{hayward2006formation,bronnikov2001regular, dymnikova1992vacuum,dymnikova2003spherically, ayon1998regular, ayon1999new, mars1996models}, among which we can mention the Frolov metric \cite{frolov2016notes}, which depends on the 
 length
scale parameter,
$l$, that regulates the central singularity through
a specific coupling between the electromagnetic and gravitational fields. It correpsponds to a generalization 
 of the Hayward regular metric \cite{hayward2006formation} through the inclusion of an  electric charge. 
      In Frolov's solution, the mass varies with the radius in order to smooth the central region, preventing the occurrence of infinite curvatures \cite{frolov2016notes,kala2025null}. This model is especially useful because it maintains the geometry of the horizon and various thermodynamic characteristics associated with classical black holes, while eliminating the singularity that conflicts with quantum principles \cite{kala2025null}.

The regular black hole solutions and the interesting consequences arising from these solutions have inspired further investigations
related to such black holes, as, for example, those regarding particle geodesics \cite{abbas2014geodesic,chiba2017note,zhou2012geodesic,tesefrancinaldo2024,gohain2025frolov,nascimento2025some11},
thermodynamics \cite{tesefrancinaldo2024, gohain2025frolov,nascimento2024some,saleh2018thermodynamics,molina2021thermodynamics,nascimento2024black,nascimento2024some111} and quasi-normal modes \cite{song2024quasinormal, gohain2025frolov,fernando2012quasinormal, flachi2013quasinormal, lin2013quasinormal}.

General solutions for Einstein's equations 
with spherical, flat, and cylindrical symmetries were obtained, considering a cloud of strings as the source of the gravitational field\cite{letelier1979clouds}.
In the case of spherical symmetry, when a cloud of radial strings surrounds the gravitating body, the solution obtained is, in essence, the Schwarzschild metric with subtle modification: the metric profile approximates Schwarzschild, but has a solid angle deficit that depends on the parameter linked to the presence of the cloud of strings. As a result, gravitational effects of global origin induced by  the cloud of strings appear. An example of this effect is the enlargement of the event horizon radius in relation to the Schwarzschild radius. The effect of a string cloud can also be obtained by a nonlinear gauge
theory \cite{ovgun2021black}, or considering a solid angle deficit \cite{barriola1989gravitational}.
Given the potential relevance to astrophysics, it is important to explore the gravitational consequences of a black hole immersed in a cloud of strings \cite{ahmed2025geometric}.

The thermodynamics associated with black holes represents one of the deepest connections between gravitation and quantum mechanics. 
Since the works of Bekenstein \cite{bekenstein1973black} and Hawking \cite{hawking1974black,hawking1975particle,hawking1976black},
it has been known that these objects emit thermal radiation whose temperature is proportional to surface gravity. Thus, the thermodynamic properties of black holes have been the subject of intensive study. The thermodynamic laws governing black holes, developed by Bardeen, Carter, and Hawking \cite{bardeen1973four}, create direct analogies between the physics of black holes and traditional thermodynamics, associating mass, horizon area, and surface gravity with energy, entropy, and temperature, respectively.

This study explores how a cloud of strings influences the Frolov black hole, focusing on the possible presence of singularities and energy conditions. We also examine geodesic motion and the corresponding effective potential, alongside certain thermodynamic quantities, in comparison with the standard Frolov black hole without the cloud of strings.

The structure of the article is as follows. Section \ref{sec2} presents a review of the Frolov solution and derives a Frolov modified metric in the presence of a surrounding cloud of strings. We also analyze singularities based on the calculation of the Kretschmann scalar and the study of geodesics and effective potential. Section \ref{sec3} analyzes the energy conditions for the new metric obtained. Section \ref{sec4} addresses various thermodynamic aspects, highlighting the influence of the string cloud parameter. Finally, section \ref{sec5} provides a concise summary of the conclusions.

\section{The metric}
\label{sec2}

In this section, we obtain the metric corresponding to the Frolov black hole with a cloud of strings, calculate and examine  the behavior of the Kretschmann scalar, 
investigate the effective potential and analyze the geodesics with respect to their completeness.

\subsection{The Frolov spacetime}

The black hole solution
proposed by Frolov\cite{frolov2016notes}
corresponds to a generalization of the Hayward black hole\cite{hayward2006formation}, in the sense that it contains an additional  
charge parameter.

The expression of the metric for the regular black hole obtained by Frolov\cite{frolov2016notes} is given by:

\begin{equation} \label{hayward}
    ds^2 = f(r) dt^2 - f(r)^{-1} dr^2 - r^2d\theta^2-r^2 \sin^2 \theta d\phi^2,
\end{equation}

\noindent with the function $f(r)$ given by

\begin{equation}
   f(r) = 1-\frac{r^2 \left(2 m r-Q^2\right)}{l^2 \left(2 m r+Q^2\right)+r^4}.
\label{Frolov}
\end{equation}

\noindent Note that $Q$ admits an interpretation in terms of the electric charge and in the limit $l\rightarrow 0$, this metric reproduces the Reissner-Nordström metric \cite{reissner1916eigengravitation}. 
The Frolov black hole is characterized by the cosmological constant $\Lambda=3/l^2$, where $l$ is the Hubble length (length scale parameter).\cite{song2024quasinormal,hayward2006formation}

Let us now analyze the regularity of Frolov's black hole by considering the behavior of the Kretschmann scalar, given by

\begin{equation}
\label{eq:2.40.1}
\begin{split}
K &= R_{\alpha\beta\mu\nu}R^{\alpha\beta\mu\nu}\\
&=+\frac{192 l^2 m^2 Q^2 r^{11} \left(15 l^2 m+2 r^3\right)}{\left(l^2 \left(2 m r+Q^2\right)+r^4\right)^6}
\\
&+\frac{8 Q^2 r^{16} \left(7 Q^2-12 m r\right)}{\left(l^2 \left(2 m r+Q^2\right)+r^4\right)^6}+\frac{24 l^8 Q^{12}}{\left(l^2 \left(2 m r+Q^2\right)+r^4\right)^6}
\\
&-\frac{32 l^8 m Q^8 r \left(7 m r+Q^2\right)}{\left(l^2 \left(2 m r+Q^2\right)+r^4\right)^6}+\frac{32 l^6 m Q^2 r^5 \left(96 l^2 m^4+19 Q^6\right)}{\left(l^2 \left(2 m r+Q^2\right)+r^4\right)^6}
\\
&+\frac{80 l^6 Q^4 r^4 \left(24 l^2 m^4-Q^6\right)}{\left(l^2 \left(2 m r+Q^2\right)+r^4\right)^6}+\frac{16 l^4 Q^2 r^8 \left(37 Q^6-120 l^2 m^4\right)}{\left(l^2 \left(2 m r+Q^2\right)+r^4\right)^6}
\\
&-\frac{16 l^2 m Q^4 r^7 \left(76 l^4 m^2+131 l^2 m r^3-34 r^6\right)}{\left(l^2 \left(2 m r+Q^2\right)+r^4\right)^6}
\\
&+\frac{16 l^2 Q^6 r^3 \left(8 l^6 m^3+38 l^4 m^2 r^3-80 l^2 m r^6-17 r^9\right)}{\left(l^2 \left(2 m r+Q^2\right)+r^4\right)^6}.
\end{split}
\end{equation}

In the limit $r \rightarrow 0$, we get

\begin{equation}
\begin{aligned}
\lim_{r\rightarrow 0} K&=\frac{24}{l^4},
\end{aligned}
\label{eq:2.40.2}
\end{equation}

\noindent which means, according to the criteria to search for singularities based on the behaviorof the Kretschmann scalar that 
the
Frolov 
black hole  is regular at the origin.



\subsection{Frolov black hole with a cloud of strings}

For a spacetime that is isotropic, static, and spherically symmetric, the line element may be represented as follows \cite{d2022introducing}:

\begin{equation}
ds^2=e^\nu dt^2-e^\lambda dr^2-r^2 d\theta^2-r^2\sin^2\theta d\phi^2.
\label{frolovcseq:10.1}
\end{equation}

For this metric tensor, the only non-null components of the Einstein tensor 
are the following
:

\begin{equation}
G_{t}^{\;t}=e^{-\lambda}\left(\frac{\lambda'}{r}-\frac{1}{r^2}\right)+\frac{1}{r^2},
\label{frolovcseq:10.2}
\end{equation}

\begin{equation}
G_{r}^{\;r}=-e^{-\lambda}\left(\frac{\nu'}{r}+\frac{1}{r^2}\right)+\frac{1}{r^2},
\label{frolovcseq:10.3}
\end{equation}

\begin{equation}
G_{\theta}^{\;\theta}=G_{\phi}^{\;\phi}=\frac{1}{2}e^{-\lambda}\left(\frac{\nu'\lambda'}{2}+\frac{\lambda'}{r}-\frac{\nu'}{r}-\frac{\nu'^2}{2}-\nu''\right).
\label{frolovcseq:10.4}
\end{equation}

%

In order obtain the Frolov black hole surrounded by a cloud o strings,
let us 
assume that the total energy-momentum tensor is given by a linear superposition of 
the energy-momentun
tensors,
$T_{\mu\nu}^{(F)}$ and $T_{\mu\nu}^{(CS)}$, where the first refers to the source of the original Frolov'sblack hole and the second is related to the cloud of strings.
It is worth pointing out that  we will assume that these tensors do not interact with each other. Thus, the total energy-momentum tensor is simply given by the linear superposition of the corresponding tensors as:

\begin{equation}
T_{\mu\nu}=T_{\mu\nu}^{(F)}+T_{\mu\nu}^{(CS)}.
\label{frolovcseq:10.5}
\end{equation}

%

Now, using the function $f(r)$ given by Eq. (\ref{Frolov}), the components of the  Einstein tensor  can be written as \cite{nascimento2025some11}:

\begin{equation}
G_t^{\;t}=G_r^{\;r}=\frac{l^2 \left(12 m^2 r^2+4 m Q^2 r-3 Q^4\right)+Q^2 r^4}{\left(l^2 \left(2 m r+Q^2\right)+r^4\right)^2},
\label{frolovcseq:1.18}
\end{equation}
\begin{equation}
\begin{aligned}
G_\theta^{\;\theta}=G_\phi^{\;\phi}=&\frac{24 l^2 m^2 r^3 \left(l^2 m-r^3\right)}{\left(l^2 \left(2 m r+Q^2\right)+r^4\right)^3}+\frac{6 l^2 Q^4 r \left(l^2 m+2 r^3\right)}{\left(l^2 \left(2 m r+Q^2\right)+r^4\right)^3}
\\
&-\frac{Q^2 \left(l^4 \left(3 Q^4-28 m^2 r^2\right)+6 l^2 m r^5+r^8\right)}{\left(l^2 \left(2 m r+Q^2\right)+r^4\right)^3},
\end{aligned}
\label{frolovcseq:1.19}
\end{equation}

\noindent which, according to Einstein's equations, are proportional to the corresponding components of the energy-momentum tensor of the source.

%
%
A particle whose four-velocity is $u^{\mu}=dx^\mu/d\lambda$ (with $\lambda$ being an independent parameter) has a world line given by $x=x(\lambda)$. By contrast, if we consider an infinitesimally thin string moving instead of a particle, the trajectory is described by a two-dimensional world sheet $\Sigma$, which can be derived by \cite{letelier1979clouds}

\begin{equation}
x^{\mu}=x^{\mu}(\lambda^{a}), \; \;a=0,1.
\label{eq:1.10}
\end{equation}
Here, $\lambda_0$ and $\lambda_1$ denote timelike and spacelike parameters, respectively.
Consequently, rather than the four-velocity $u^{\mu}$, we encounter a bivector $\Sigma^{\mu\nu}$, defined by \cite{letelier1979clouds}:

\begin{equation}
\Sigma^{\mu\nu}=\epsilon^{ab}\frac{\partial{x^\mu}}{\partial{\lambda^{a}}}\frac{\partial{x^\nu}}{\partial{\lambda^{b}}}.
\label{eq:1.11}
\end{equation}

On this world sheet, there will be an induced metric, $\gamma_{ab}$, with $a, b = 0, 1$, such that,

\begin{equation}
\gamma_{ab}=g_{\mu\nu}\frac{\partial{x^\mu}}{\partial{\lambda^{a}}}\frac{\partial{x^\nu}}{\partial{\lambda^{b}}},
\label{eq:1.12}
\end{equation}
whose determinant is given by $\gamma$.
For a cloud of strings, we have \cite{letelier1979clouds}

\begin{equation}
T^{\mu\nu}=\rho\frac{\Sigma^{\mu\beta}\Sigma_{\beta}^{\;\nu}}{(-\gamma)^{1/2}},
\label{frolovcseq:10.8}
\end{equation}

\noindent where $\gamma=\frac{1}{2}\Sigma^{\mu\nu}\Sigma_{\mu\nu}$.

The non-null components of the energy-momentum tensor for a spherically symmetric cloud of strings are given by \cite{letelier1979clouds}:

\begin{equation}
T_{0}^{\;0}=T_{1}^{\;1}=\frac{a}{r^2},
\label{frolovcseq:10.10}
\end{equation}

\begin{equation}
T_{2}^{\;2}=T_{3}^{\;3}=0.
\label{frolovcseq:10.11}
\end{equation}


Using Eqs. (\ref{frolovcseq:10.2})-(\ref{frolovcseq:10.4}) for the components of the Einstein tensor together with Eqs. (\ref{frolovcseq:1.18})-(\ref{frolovcseq:1.19}) and (\ref{frolovcseq:10.10})-(\ref{frolovcseq:10.11}) for the energy-momentum tensors for Frolov space-time and the cloud of strings, respectively, we get the following equations:

\begin{equation}
e^{-\lambda}\left(\frac{\lambda'}{r}-\frac{1}{r^2}\right)+\frac{1}{r^2}=\frac{a}{r^2}+\frac{l^2 \left(12 m^2 r^2+4 m Q^2 r-3 Q^4\right)+Q^2 r^4}{\left(l^2 \left(2 m r+Q^2\right)+r^4\right)^2},
\label{frolovcseq:10.16}
\end{equation}

\begin{equation}
-e^{-\lambda}\left(\frac{\nu'}{r}+\frac{1}{r^2}\right)+\frac{1}{r^2}=\frac{a}{r^2}+\frac{l^2 \left(12 m^2 r^2+4 m Q^2 r-3 Q^4\right)+Q^2 r^4}{\left(l^2 \left(2 m r+Q^2\right)+r^4\right)^2},
\label{frolovcseq:10.17}
\end{equation}

\begin{equation}
\begin{aligned} 
\frac{1}{2}e^{-\lambda}\left(\frac{\nu'\lambda'}{2}+\frac{\lambda'}{r}-\frac{\nu'}{r}-\frac{\nu'^2}{2}-\nu''\right)=& +\frac{24 l^2 m^2 r^3 \left(l^2 m-r^3\right)}{\left(l^2 \left(2 m r+Q^2\right)+r^4\right)^3}
\\
&+\frac{6 l^2 Q^4 r \left(l^2 m+2 r^3\right)}{\left(l^2 \left(2 m r+Q^2\right)+r^4\right)^3}
\\
&-\frac{Q^2 \left(l^4 \left(3 Q^4-28 m^2 r^2\right)+6 l^2 m r^5+r^8\right)}{\left(l^2 \left(2 m r+Q^2\right)+r^4\right)^3}.
\end{aligned}
\label{frolovcseq:10.18}
\end{equation}

\noindent By subtracting Eqs.(\ref{frolovcseq:10.16}) and (\ref{frolovcseq:10.17}) we arrive at:

\begin{equation}
\lambda=-\nu\Rightarrow\lambda'=-\nu'.
\label{frolovcseq:10.19}
\end{equation}

Adding Eqs. (\ref{frolovcseq:10.16}) and (\ref{frolovcseq:10.17}) and using the relation expressed by  Eq. (\ref{frolovcseq:10.19}), we find the following result

\begin{equation}
e^{-\lambda}\frac{\lambda'}{r}-e^{-\lambda}\frac{1}{r^2}+\frac{1}{r^2}=\frac{a}{r^2}+\frac{l^2 \left(12 m^2 r^2+4 m Q^2 r-3 Q^4\right)+Q^2 r^4}{\left(l^2 \left(2 m r+Q^2\right)+r^4\right)^2}.
\label{frolovcseq:10.20}
\end{equation}

Using the relation that follows, 

\begin{equation}
\nu=-\lambda=ln(1+f(r)),
\label{frolovcseq:10.21}
\end{equation}

\noindent and considering Eqs. (\ref{frolovcseq:10.19}) and (\ref{frolovcseq:10.21}), we can rewrite Eqs. (\ref{frolovcseq:10.20}) and (\ref{frolovcseq:10.18}), respectively, as 

\begin{equation}
-\frac{1}{r^2}(rf'+f)=\frac{a}{r^2}+\frac{l^2 \left(12 m^2 r^2+4 m Q^2 r-3 Q^4\right)+Q^2 r^4}{\left(l^2 \left(2 m r+Q^2\right)+r^4\right)^2},
\label{frolovcseq:10.22}
\end{equation}

\begin{equation}
\begin{aligned}
2\frac{f'}{r}+f''=& -2\frac{24 l^2 m^2 r^3 \left(l^2 m-r^3\right)}{\left(l^2 \left(2 m r+Q^2\right)+r^4\right)^3}
\\
&-2\frac{6 l^2 Q^4 r \left(l^2 m+2 r^3\right)}{\left(l^2 \left(2 m r+Q^2\right)+r^4\right)^3}
\\
&+2\frac{Q^2 \left(l^4 \left(3 Q^4-28 m^2 r^2\right)+6 l^2 m r^5+r^8\right)}{\left(l^2 \left(2 m r+Q^2\right)+r^4\right)^3}.
\end{aligned}
\label{frolovcseq:10.23}
\end{equation}

Adding Eqs. (\ref{frolovcseq:10.22}) and (\ref{frolovcseq:10.23}) and multiplying the result by $r^2$, we get the following differential equation:

\begin{equation}
\begin{aligned}
&r^2f''+rf'-f-a-\frac{l^2 r^2\left(12 m^2 r^2+4 m Q^2 r-3 Q^4\right)+Q^2 r^6}{\left(l^2 \left(2 m r+Q^2\right)+r^4\right)^2}
\\
& +\frac{48 l^2 m^2 r^5 \left(l^2 m-r^3\right)}{\left(l^2 \left(2 m r+Q^2\right)+r^4\right)^3}+\frac{12 l^2 Q^4 r^3 \left(l^2 m+2 r^3\right)}{\left(l^2 \left(2 m r+Q^2\right)+r^4\right)^3}
\\
&-\frac{2Q^2 r^2\left(l^4 \left(3 Q^4-28 m^2 r^2\right)+6 l^2 m r^5+r^8\right)}{\left(l^2 \left(2 m r+Q^2\right)+r^4\right)^3}=0.    
\end{aligned}
\label{frolovcseq:10.24}
\end{equation}

\noindent The general solution of the above differential equation is given by:

\begin{equation}
f(r)=\frac{-2 a l^2 m r^2-a l^2 Q^2 r-a r^5+4 l^2 m^2 r+2 l^2 m Q^2+Q^2 r^3}{r \left(2 l^2 m r+l^2 Q^2+r^4\right)}+\frac{C_1}{r}+C_2 r.
\label{frolovcseq:10.25}
\end{equation}

\noindent 

Assuming, for convenience, that  $C_1=-2m$ and $C_2=0$ and performing the simplifications,
Eq.(\ref{frolovcseq:10.25}) can be written as

\begin{equation}
f(r)=-a-\frac{\left(2 m r-Q^2\right)r^2}{r^4+\left(2 m r+Q^2\right)l^2}.
\label{frolovcseq:10.26}
\end{equation}

Thus, substituting Eq. (\ref{frolovcseq:10.26}) into Eq. (\ref{frolovcseq:10.21}), we obtain:

\begin{equation}
\nu=-\lambda=ln\left[1-a-\frac{\left(2 m r-Q^2\right)r^2}{r^4+\left(2 m r+Q^2\right)l^2}\right].
\label{frolovcseq:10.27}
\end{equation}

Substituting Eq. (\ref{frolovcseq:10.27}) into Eq. (\ref{frolovcseq:10.1}), we finally obtain the Frolov metric for a black hole with a cloud of strings, which is given by

\begin{equation}
 \begin{aligned}
ds^2=&\left(1-a-\frac{\left(2 m r-Q^2\right)r^2}{r^4+\left(2 m r+Q^2\right)l^2}\right)dt^2\\
&-\left(1-a-\frac{\left(2 m r-Q^2\right)r^2}{r^4+\left(2 m r+Q^2\right)l^2}\right)^{-1}dr^2-r^2 d\Omega^2.
\label{frolovcseq:10.28}
\end{aligned} 
\end{equation}

Note that if $Q=0$, we get the Hayward metric with cloud of string \cite{nascimento2024some}. If $l=0$, the metric (\ref{frolovcseq:10.28}) becomes similar to the Reissner-Nordström spacetime.


\subsection{Analysis of the Kretschmann scalar}

The divergence of the Kretschmann scalar does not imply that there is no complete geodesic, nor geodesic incompleteness necessarily implies the existence of a physical singularity. Thus, we have to combine the analysis of the scalar invariant constructed with the curvature and the analysis of the geodesics. 

For the metric given by Eq. (\ref{frolovcseq:10.28}), the Kretschmann scalar,
$K=R_{\alpha\beta\mu\nu}R^{\alpha\beta\mu\nu}$,
is given by

\begin{equation}
\begin{aligned}
K=
&=+\frac{4 a^2}{r^4}-\frac{16 l^6 Q^{10} r \left(2 l^2 m+5 r^3\right)}{\left(l^2 \left(2 m r+Q^2\right)+r^4\right)^6}+\frac{32 l^6 m Q^8 r^2 \left(19 r^3-7 l^2 m\right)}{\left(l^2 \left(2 m r+Q^2\right)+r^4\right)^6}
\\
&+\frac{16 l^2 Q^6 r^3 \left(8 l^6 m^3+38 l^4 m^2 r^3-80 l^2 m r^6-17 r^9\right)}{\left(l^2 \left(2 m r+Q^2\right)+r^4\right)^6}
\\
&+\frac{16 l^2 m Q^4 r^4 \left(120 l^6 m^3-76 l^4 m^2 r^3-131 l^2 m r^6+34 r^9\right)}{\left(l^2 \left(2 m r+Q^2\right)+r^4\right)^6}
\\
&+\frac{96 m Q^2 r^5 \left(32 l^8 m^4-20 l^6 m^3 r^3+30 l^4 m^2 r^6+4 l^2 m r^9-r^{12}\right)}{\left(l^2 \left(2 m r+Q^2\right)+r^4\right)^6}
\\
&+\frac{48 m^2 r^6 \left(32 l^8 m^4-16 l^6 m^3 r^3+72 l^4 m^2 r^6-8 l^2 m r^9+r^{12}\right)}{\left(l^2 \left(2 m r+Q^2\right)+r^4\right)^6}
\\
&+\frac{8 Q^4 \left(3 l^8 Q^8+74 l^4 Q^4 r^8+7 r^{16}\right)}{\left(l^2 \left(2 m r+Q^2\right)+r^4\right)^6}-\frac{8 a \left(Q^2-2 m r\right)}{l^2 r^2 \left(2 m r+Q^2\right)+r^6},
\label{frolovcs1}
\end{aligned}
\end{equation}

\noindent  whose limits as $r\rightarrow 0$ and $r\rightarrow \infty$, are the following

\begin{equation}
\lim_{r\rightarrow 0}K=\infty,
\label{frolovcs2}
\end{equation}
\begin{equation}
\lim_{r\rightarrow \infty}K=0.
\label{frolovcs3}
\end{equation}

Note that the Kretschmann scalar tends to infinity as 
as $ r\rightarrow 0$, differently from the case of the original
Frolov's
black hole solution.
Thus, we can conclude that the addition of a cloud of strings alters the metric by breaking its regularity, or in other words, the Frolov black hole metric with a cloud of strings is singular at the origin ($r=0$).

\subsection{Analysis of geodesics and effective potential}

Now consider the spherically symmetric and static solution, Eq. (\ref{frolovcseq:10.1}), with $f(r)$ given by:

\begin{equation}
 \begin{aligned}
f(r)=1-a-\frac{\left(2 m r-Q^2\right)r^2}{r^4+\left(2 m r+Q^2\right)l^2}.
\label{eq:frolovcs}
\end{aligned} 
\end{equation}

Using the geodesic action together with the variational principle, we formulate a Lagrangian in which the metric is described as

\begin{equation}
    2\mathcal{L} = f(r) \dot{t}^2-\frac{1}{f(r)}\dot{r}^2-r^2\dot{\theta}^2-r^2\sin^2\theta\dot{\phi}^2,\label{lagrangian}
\end{equation}
where the 'point' represents the derivative in relation to proper time, $\tau$.
Restricting our analysis of geodesics to the equatorial plane of the black hole and assuming the Euler-Lagrange equations, we obtain that

\begin{align}
    E &= f(r)\dot{t},
\label{energia}
\end{align}
\begin{align}
    J &= -r^2\dot{\phi},
\label{momento angular}
\end{align}

\noindent where $E$ and $J$ are movement constants that can be interpreted as energy and angular momentum of the particle that is moving near the black hole. Substituting Eqs. (\ref{energia}) and (\ref{momento angular}) in Eq. (\ref{lagrangian}), we arrive at:

\begin{equation}
\dot{r}^2=E^2-V_{eff},
\label{energia e potencial efetivo}
\end{equation}

\noindent where

\begin{equation}
V_{eff}=f(r)\left(\frac{J^2}{r^2}+L\right).
\label{potenciale efetivo}
\end{equation}

For a probe massive particle $(L=1)$ falling radially $(J=0)$ into the black hole, the radial geodesic equation of motion is:

\begin{equation}
    \dot{r}^2 = E^2- f(r),
    \label{eqmovimento}
\end{equation}
while the effective potential reads:
\begin{equation}
    V_{eff} = f(r).
\end{equation}

\noindent By applying the relationship

\begin{equation}
\begin{aligned}
  \left(\frac{dr}{dt}\right)^2\dot{t}^2=\dot{r}^2,
\end{aligned}
\label{eq:1.85.1}
\end{equation}

\noindent into Eq. (\ref{eqmovimento}) while employing Eq. (\ref{energia}), leading to

\begin{equation}
\left(\frac{dr}{dt}\right)^2=f(r)^2\left(1-\frac{f(r)}{E^2}\right).
\label{eq:1.86.1}
\end{equation}

Incorporating Eq. (\ref{eq:frolovcs}) into Eq. (\ref{eq:1.86.1}), one obtains the relation between the coordinates $t$ and $r$ governing the particle’s radial motion:

\begin{equation}
\pm t=\int\frac{dr}{\sqrt{\left(1-a-\frac{\left(2 m r-Q^2\right)r^2}{r^4+\left(2 m r+Q^2\right)l^2}\right)^2-\frac{\left(1-a-\frac{\left(2 m r-Q^2\right)r^2}{r^4+\left(2 m r+Q^2\right)l^2}\right)^3}{E^2}}}.
\label{eq:1.91.1}
\end{equation}

Using Eq.(\ref{eqmovimento}), one derives the relation between the proper time $\tau$ and the radial coordinate $r$:

\begin{equation*}
\left(\frac{dr}{d\tau}\right)^2=E^2-f(r),
\end{equation*}

\begin{equation}
\pm\tau=\int\frac{dr}{\sqrt{E^2-\left(1-a-\frac{\left(2 m r-Q^2\right)r^2}{r^4+\left(2 m r+Q^2\right)l^2}\right)}}.
\label{eq:1.92.1}
\end{equation}

To assess the regularity of the solution, we must evaluate the integral in Eq. (\ref{eq:1.92.1}) along a geodesic that approaches $r=0$. If we take $r \approx 0$, we obtain that:

\begin{equation}
\pm\tau \approx \int \left[\frac{1}{\sqrt{a+E^2-1}}+\frac{r^2}{2 l^2 \left(a+E^2-1\right)^{3/2}}\right]dr,
\label{eq:1.92.3}
\end{equation}

\noindent whose approximate solution is 
\begin{equation}
\pm\tau \approx \frac{6 l^2 r \left(a+E^2-1\right)+r^3}{6 l^2 \left(a+E^2-1\right)^{3/2}}.
\label{eq:1.92.3.1}
\end{equation}

Thus, if we consider $r\approx0$, for $E^2>1-a$, the proper time is finite, and as a consequence, the geodesics are complete. 

\begin{figure}
    \centering
    \includegraphics[scale=0.49]{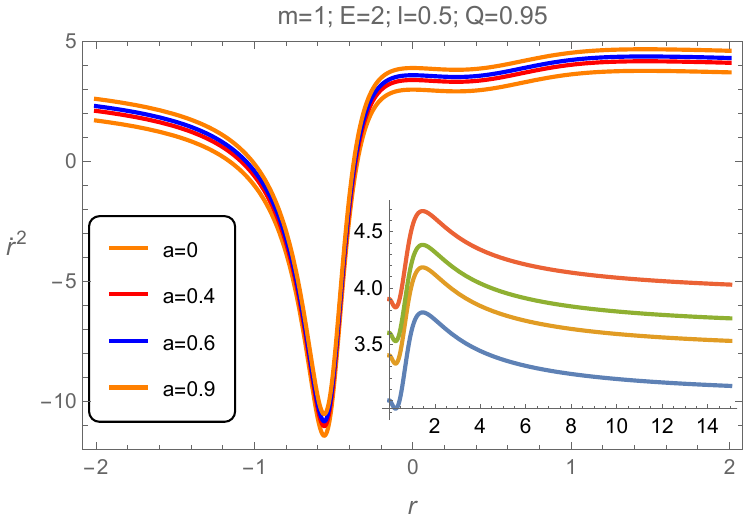}
    \label{im1afrolov} 
    \includegraphics[scale=0.49]{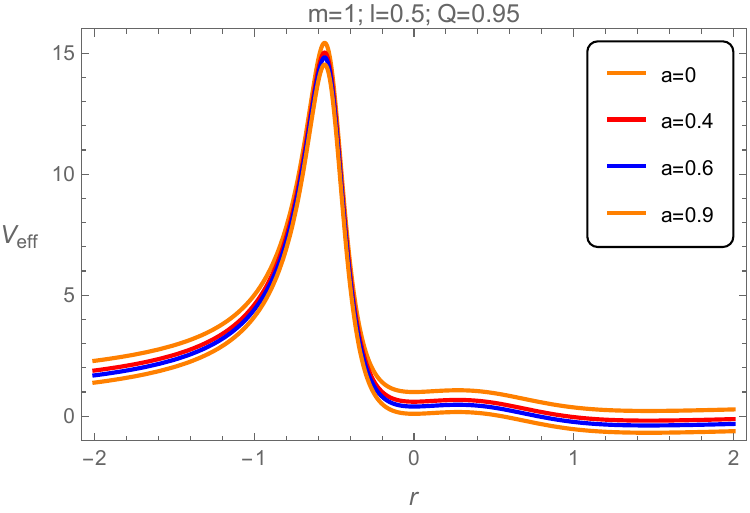}
    \label{im1bfrolov} 
    \includegraphics[scale=0.49]{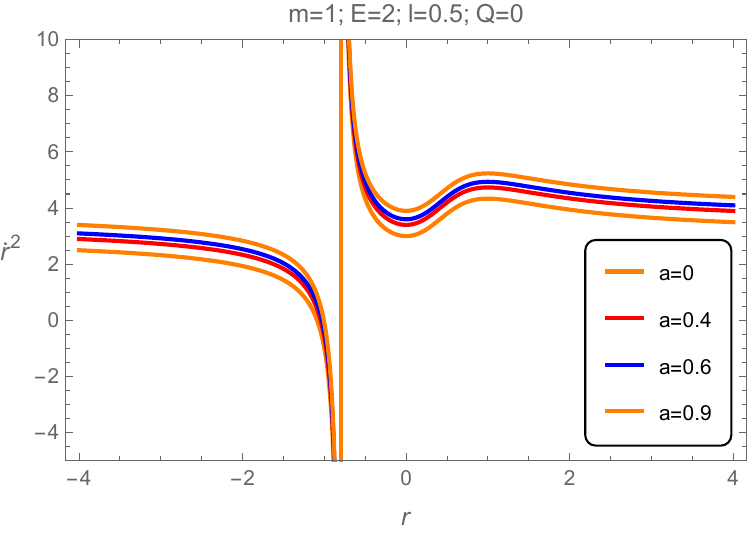}
    \label{im1cfrolov} 
    \includegraphics[scale=0.49]{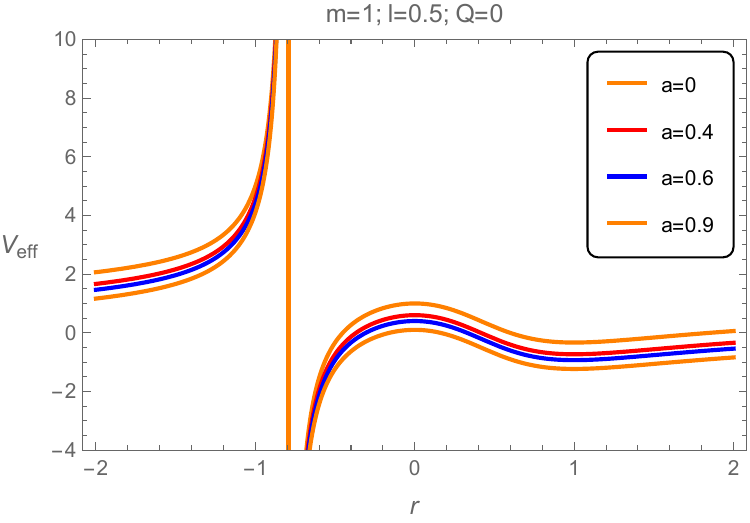}
    \label{im1dfrolov}
    \includegraphics[scale=0.49]{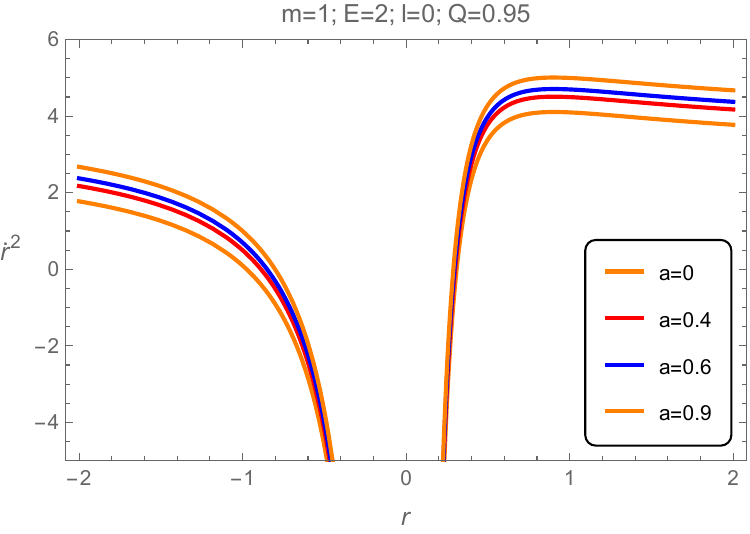}
    \label{im1efrolov}
    \includegraphics[scale=0.49]{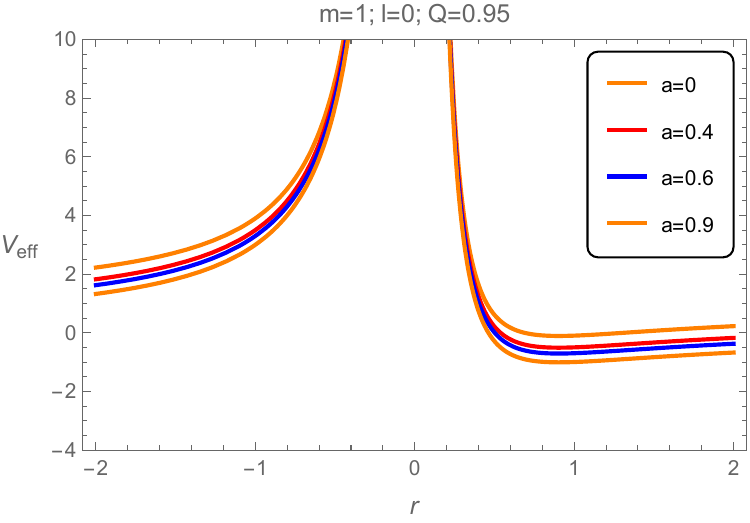}
    \label{im1ffrolov}
    \caption{The left plot shows $\dot{r}^2$ for $E=2$. The right plot shows the effective potential $V_{eff} = f(r)$.}
    \label{testandofrolovcs}
  \end{figure}

Now, let us compare the results predicted by the values of the Kretschmann scalar when $r\rightarrow 0$, with the features of the geodesics, namely if they are complete or incomplete. The plots for $\dot{r}^2$ and $V_{eff}$ are shown in Fig. (\ref{testandofrolovcs}).

For the Frolov spacetime with cloud of strings $(0<a<1)$ and without cloud of strings $(a=0)$, the test particle manages to cross the potential barrier and reach the point $r=0$ in a finite time. This indicates that the geodesics are complete, and therefore the Frolov spacetime and the Frolov spacetime with cloud of strings are regular.  

This conclusion for Frolov spacetime is confirmed by the fact that the Kretschmann scalar is finite when $r\rightarrow 0$, Eq. (\ref{eq:2.40.2}). However, for the Frolov spacetime with a cloud of strings, the analysis of the geodesics suggests a regular spacetime at the origin, while the Kretschmann scalar diverges when $r\rightarrow 0$, Eq. (\ref{frolovcs2}), suggesting a singular metric at the origin.

Note that, for $l>0$, $Q=0$ and $(0<a<1)$, which represents Hayward space-time with cloud of strings and without $(a=0)$ cloud of strings, we can conclude that the particle reaches the point $r=0$ in a finite time. This indicates that the geodesics are complete and therefore space-time is regular at the origin. This conclusion about the Hayward black hole with cloud of strings is not confirmed by the Kretschmann scalar, which is infinite when $r\rightarrow 0$ \cite{nascimento2024some}.

In the Reissner-Nordström spacetime, we can conclude that the particle cannot reach the point $r=0$ in a finite time. This indicates that the geodesics are incomplete, and therefore the spacetime is singular at the origin. This conclusion is confirmed by the fact that the Kretschmann scalar is infinite when $r\rightarrow 0$.

In summary, the analysis of geodesics, aimed at knowing the characteristics of space-time in relation to the singularity, can confirm or not what is predicted by the results provided by the Kretschmann scalar. Thus, in order to effectively verify the existence or not of singularities in space-time, an analysis of geodesic completeness or incompleteness is necessary, since the analysis of the Kretschmann scalar only gives us a local view of possible singularities in the metric.

 \section{Energy conditions}
 \label{sec3}

Within General Relativity, energy conditions were proposed as physically reasonable restrictions on matter. The four principal conditions are the weak (WEC), strong (SEC), dominant (DEC), and null (NEC) energy conditions
\cite{kontou2020energy}.

It is understood that the energy conditions are violated by regular black holes \cite{zaslavskii2010regular}. With the black hole metric becoming singular after incorporating a cloud of strings, it is important to examine the energy conditions of the resulting metric.

To examine the energy conditions, we must determine the components of the stress-energy tensor, which are listed below:

\begin{equation}
    T^\mu_{\mu} = \text{diag}[\rho, -p_r, -p_t, -p_t],
\end{equation}

\noindent where $\rho$ is the energy density, $p_r$ is the radial pressure, and $p_t$ is the tangential pressure.
Considering the Einstein
equations and Eq. (\ref{frolovcseq:10.28}), we find

\begin{align}
    \rho = \frac{1 - f(r) - r f'(r)}{\kappa^2 r^2}, \\
    p_r = \frac{r f'(r) + f - 1}{\kappa^2 r^2},\\
    p_t = \frac{rf''(r) + 2 f'(r)}{2 \kappa^2 r},
\end{align}

with the function $f(r)$ defined as

\begin{equation}
f(r)=1-a-\frac{\left(2 m r-Q^2\right)r^2}{r^4+\left(2 m r+Q^2\right)l^2}.
\label{eq:1.86.1.1}
\end{equation} 

The conditions can be expressed through the following inequalities \cite{rodrigues2022embedding,kontou2020energy}:

\begin{align}
    NEC_{1,2} = WEC_{1,2} = SEC_{1,2} &\Leftrightarrow \rho + p_{r,t} \geq 0,\\
    SEC_3 &\Leftrightarrow \rho +p_r + 2p_t \geq 0,\\
    DEC_{1,2} &\Leftrightarrow \rho -|p_{r,t}| \geq 0,\\
    DEC_3 = WEC_3 &\Leftrightarrow \rho \geq 0,
\end{align}
where the indexes $1$ and $2$ refer, respectively, to the radial and tangential components of the pressure. We see that  $DEC_{1,2}\Leftrightarrow ((NEC_{1,2})$ and $(\rho-p_{r,t}\geq 0))$, so, we replace $DEC_{1,2}\Rightarrow \rho-p_{r,t}\geq 0$.

Proceeding, assume that $f(r) > 0$, which makes $t$ a time-like coordinate; accordingly, the energy conditions take the following form:

\begin{equation}
    NEC_1 \Leftrightarrow 0,
\end{equation}

\begin{equation}
\begin{aligned}
    NEC_2& \Leftrightarrow \frac{a}{\kappa ^2 r^2}
    \\
    &+\frac{2 r \left(-4 l^4 m Q^2 \left(m r+Q^2\right)+l^2 r^3 \left(18 m^2 r^2+6 m Q^2 r-7 Q^4\right)+Q^2 r^7\right)}{\kappa ^2 \left(l^2 \left(2 m r+Q^2\right)+r^4\right)^3}\geq 0,
\end{aligned}
\end{equation}

\begin{equation}
\begin{aligned}
    WEC_3 &\Leftrightarrow \frac{a}{\kappa ^2 r^2}+\frac{l^2 \left(12 m^2 r^2+4 m Q^2 r-3 Q^4\right)+Q^2 r^4}{\kappa ^2 \left(l^2 \left(2 m r+Q^2\right)+r^4\right)^2}\geq 0,
\end{aligned}
\end{equation}

\begin{equation}
\begin{aligned}
    SEC_3 &\Leftrightarrow \frac{48 l^2 m^2 r^3 \left(r^3-l^2 m\right)}{\kappa ^2 \left(l^2 \left(2 m r+Q^2\right)+r^4\right)^3}
    \\
    &+\frac{2 Q^2 \left(l^4 \left(-28 m^2 r^2-6 m Q^2 r+3 Q^4\right)+6 l^2 r^4 \left(m r-2 Q^2\right)+r^8\right)}{\kappa ^2 \left(l^2 \left(2 m r+Q^2\right)+r^4\right)^3}\geq 0,
\end{aligned}
\end{equation}

\begin{equation}
\begin{aligned}
    DEC_1 &\Leftrightarrow \frac{2 a}{\kappa ^2 r^2}+\frac{l^2 \left(24 m^2 r^2+8 m Q^2 r-6 Q^4\right)+2 Q^2 r^4}{\kappa ^2 \left(l^2 \left(2 m r+Q^2\right)+r^4\right)^2}\geq 0,
\end{aligned}
\end{equation}

\begin{equation}
\begin{aligned}
    DEC_2 &\Leftrightarrow \frac{a}{\kappa ^2 r^2}+\frac{12 l^2 m^2 r^3 \left(4 l^2 m-r^3\right)}{\kappa ^2 \left(l^2 \left(2 m r+Q^2\right)+r^4\right)^3}
    \\
    &+\frac{l^4 \left(48 m^2 Q^2 r^2+4 m Q^4 r-6 Q^6\right)+10 l^2 Q^4 r^4}{\kappa ^2 \left(l^2 \left(2 m r+Q^2\right)+r^4\right)^3}\geq 0.
\end{aligned}
\end{equation}

Taking the limit as $r \rightarrow \infty$, we obtain:

\begin{align}
    NEC_1 &\Leftrightarrow  0, \\
    NEC_2 &\Leftrightarrow 0, \\
    WEC_3 &\Leftrightarrow 0, \\
    SEC_3 &\Leftrightarrow 0,\\
    DEC_1 &\Leftrightarrow 0,\\
    DEC_2 &\Leftrightarrow 0.
\end{align}

The findings on the energy conditions indicate that, in the scenario examined, these conditions may hold or be violated depending on the parameter relationships. For very large radial distances, the analysis shows that all conditions are satisfied and reduce to zero.
Note that if one instead takes $f(r) < 0$, making $t$ space-like, the conclusions regarding the energy conditions remain unchanged.

\section{Black hole thermodynamic framework}
\label{sec4}

\subsection{Black hole mass}

Let the horizon radius be $r_h$, since $f(r_h)=0$ with $f(r)$ as in Eq. (\ref{eq:1.86.1.1}),
one obtains a expression for the black hole mass in terms of $r_h$

\begin{equation}
m=-\frac{(a-1) r_h^4+(a-1) l^2 Q^2-Q^2 r_h^2}{2 \left((a-1) l^2 r_h+r_h^3\right)},
\label{eq:1.59frolovcordas}
\end{equation}

\noindent which is expressed using the parameter $a$, which encodes the cloud of strings. Note that if $a=0$, we recover the mass of the regular Frolov black hole, without the cloud of strings, in terms of the horizon radius. By setting $l=0$ and $a=0$ we recover the mass of the Reissner-Nordström black hole. 

Calculating the horizon area proceeds as follows:

\begin{equation}
A=\int \sqrt{-g}d\theta d\phi=4\pi r_h^{2}.
\label{eq:1.60frolovcordas}
\end{equation}

Conversely, the black hole entropy can be obtained via the area law \cite{bekenstein1973black}, employing the relation:

\begin{equation}
S=\frac{A}{4}=\pi r_h^2.
\label{eq:1.61frolovcordas}
\end{equation}

Hence, the mass parameter can be expressed as a function of the entropy

\begin{equation}
m=\frac{S \left(-a S+\pi  Q^2+S\right)-\pi ^2 (a-1) l^2 Q^2}{2 \sqrt{\pi } \sqrt{S} \left(\pi  (a-1) l^2+S\right)}.
\label{eq:1.62frolovcordas}
\end{equation}

\begin{figure}
    \centering
    \includegraphics[scale=0.49]{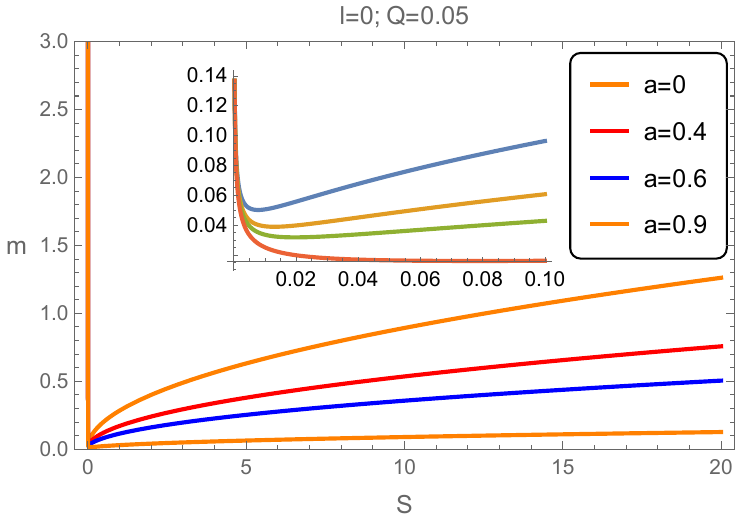}
    \label{im2afrolov} 
    \includegraphics[scale=0.49]{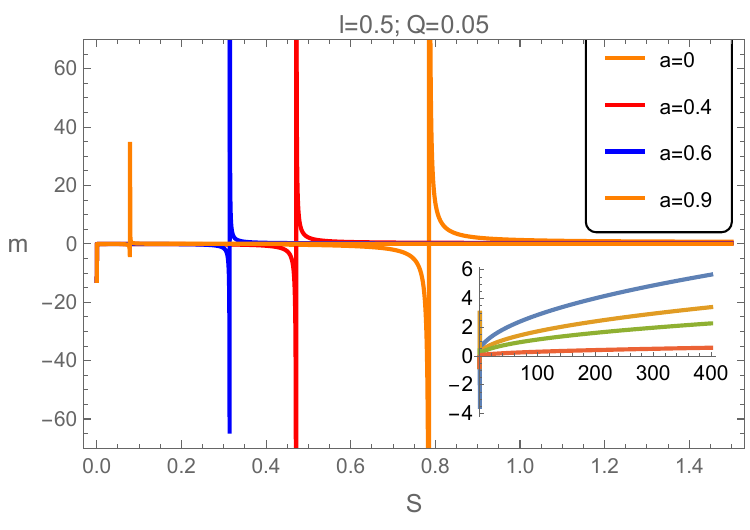}
    \label{im2bfrolov} 
    \includegraphics[scale=0.49]{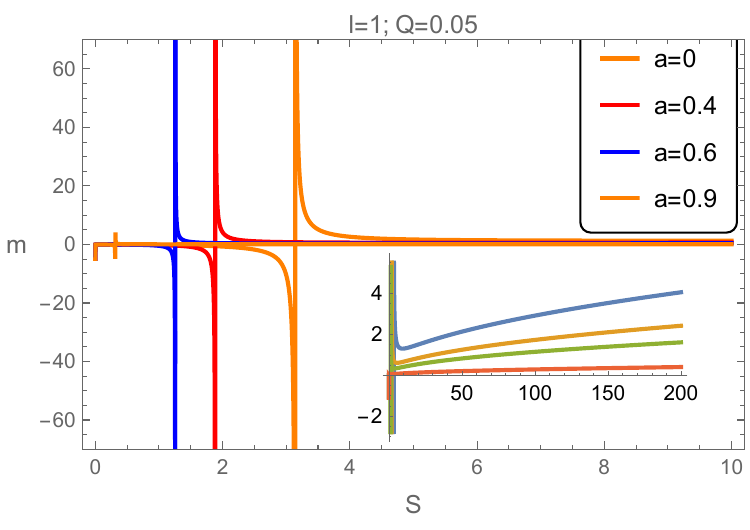}
    \label{im2cfrolov} 
    \includegraphics[scale=0.49]{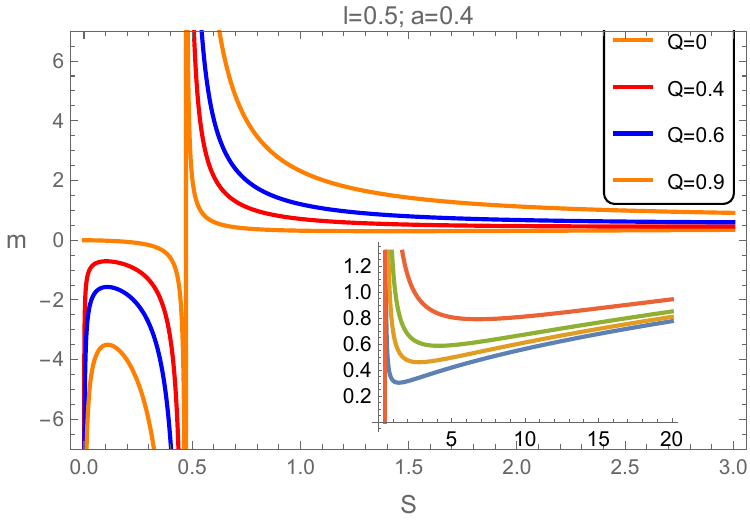}
    \label{im2dfrolov} 
    \caption{Black hole mass as a function of entropy $m(S)$ for different values of $a$, $Q$ and $l$.}
    \label{im2frolovcordas}
  \end{figure}

In Fig. \ref{im2frolovcordas}, we represent the behavior of the mass parameter, $m$, as a function of the entropy of the black hole, $S$, in different situations. Note that, for the Reissner-Nordström black hole, ($a=0$ and $l=0$), the mass parameter presents only positive values for positive values of the entropy,$S$. Similar behavior is obtained when  ($l=0$), thus we return to the Reissner-Nordström black hole scenario, but now, with a cloud of strings, $0<a<1$.

If we consider the Frolov black hole, it is possible to notice that the mass parameter has positive and negative values depending on the parameters of the black hole. This is also repeated when we consider the cloud of strings in Frolov's black hole space-time. It is important to note that the cloud of string parameter modifies the phase transition point for Frolov black hole surrounded by this cloud.


\subsection{Black hole temperature}

For the Frolov black hole in the presence of a cloud of strings, the surface gravity $(\kappa)$ is obtained from the expression below:

\begin{equation}
\kappa=\frac{f'(r)}{2}\left|\frac{}{}_{r_{h}}\right.,
\label{eq:1.63frolovcordas}
\end{equation}

\noindent Note that the symbol $'$ denotes $d/dr$. Hawking showed that black holes radiate, and in a static space-time the Hawking temperature describing this radiation is provided by \cite{hawking1975particle}:
  
\begin{equation}
T_\kappa=\frac{\kappa}{2\pi}.
\label{eq:1.64frolovcordas}
\end{equation}

Using Eq. (\ref{eq:1.64frolovcordas}) with $\kappa$ given by (\ref{eq:1.63frolovcordas}), it is possible to calculate the Hawking temperature, $T_\kappa=T$, for the Frolov black hole with the cloud of strings:

\begin{equation}
T=\frac{\pi ^3 (a-1)^2 l^4 Q^2+\pi  (a-1) l^2 S \left(4 \pi  Q^2-3 (a-1) S\right)+S^2 \left(-a S-\pi  Q^2+S\right)}{4 \sqrt{\pi } S^{3/2} \left(2 \pi ^2 l^2 Q^2+S^2\right)}.
\label{eq:1.68frolovcordas}
\end{equation}

\begin{figure}
    \centering
    \includegraphics[scale=0.49]{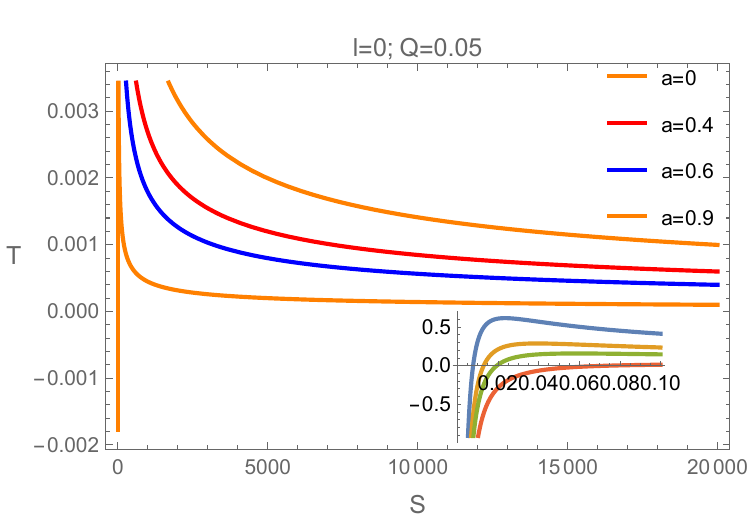}
    \label{im3afrolov} 
    \includegraphics[scale=0.49]{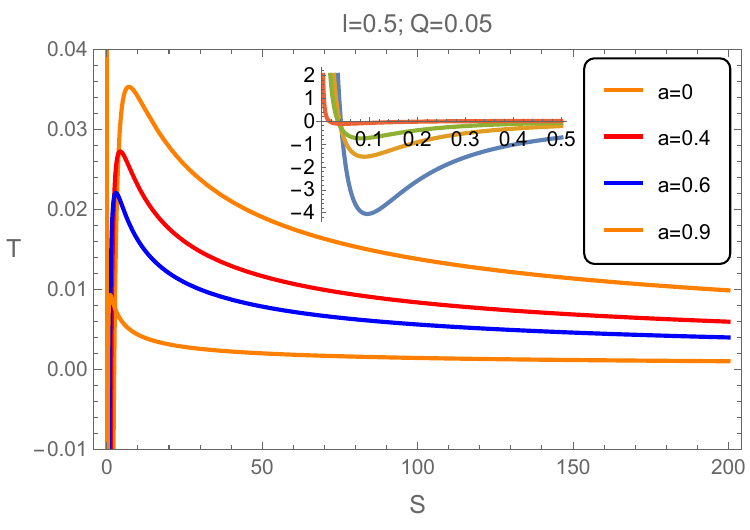}
    \label{im3bfrolov} 
    \includegraphics[scale=0.49]{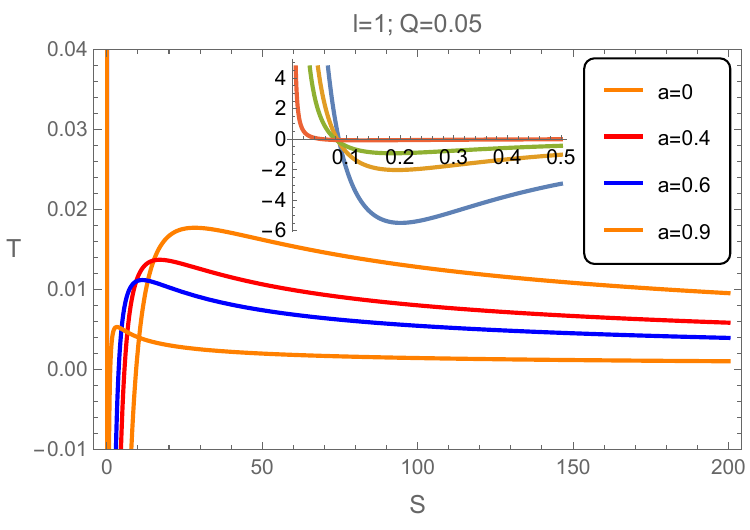}
    \label{im3cfrolov} 
    \includegraphics[scale=0.49]{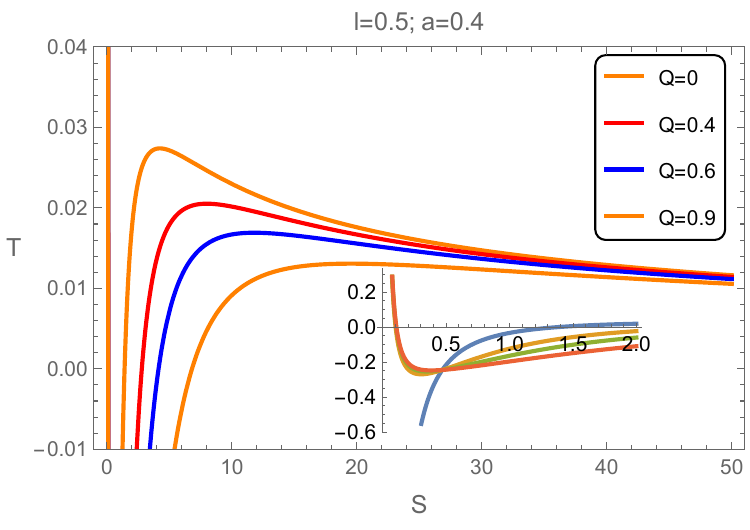}
    \label{im3dfrolov} 
    \caption{Black hole temperature as a function of entropy $T(S)$ for different values of $a$, $Q$ and $l$.}
    \label{im3frolovcordas}
  \end{figure}

In Fig. \ref{im3frolovcordas}, we represent the behavior of the temperature parameter, $T$, as a function of the entropy of the black hole, $S$, in different situations. Note that for the Reissner-Nordström space-time ($a=0$ and $l=0$) the temperature parameter can have positive or negative values for $S>0$. Analogously, for the Reissner-Nordström-Letelier spacetime ($l=0$ and $0<a<1$). Note that when we consider the Frolov spacetime ($l=1$, $Q=0.05$ and $a=0$), we can also see that the temperature parameter will have positive and negative values depending on the parameters of the black hole. This also repeats itself when we consider the cloud of strings in Frolov space-time ($l=1$, $Q=0.05$ and $0<a<1$).

\subsection{Black hole heat capacity}

To assess stability, we consider the heat capacity, which for the Frolov black hole with a cloud of strings can be computed as:

\begin{equation}
C=T\frac{\partial S}{\partial T}=T\left(\frac{\partial T}{\partial S}\right)^{-1}.
\label{eq:1.69frolovcordas}
\end{equation}

By inserting Eq.(\ref{eq:1.68frolovcordas}) into Eq.(\ref{eq:1.69frolovcordas}), we obtain the heat capacity expressed as a function of the black hole entropy:

\begin{equation}
\begin{aligned}   
C=&-[2 S \left(2 \pi ^2 l^2 Q^2+S^2\right) \left(\pi ^3 (a-1)^2 l^4 Q^2+\pi  (a-1) l^2 S \left(4 \pi  Q^2-3 (a-1) S\right)\right.
\\
&\left.+S^2 \left(-a S-\pi  Q^2+S\right)\right)]/[6 \pi ^5 (a-1)^2 l^6 Q^4+\pi ^3 (a-1) l^4 Q^2 S \left(13 (a-1) S+8 \pi  Q^2\right)
\\
&+\pi  l^2 S^2 \left(26 \pi  (a-1) Q^2 S-9 (a-1)^2 S^2+2 \pi ^2 Q^4\right)+S^4 \left(-a S-3 \pi  Q^2+S\right)]
\label{eq:1.70frolovcordas}
\end{aligned}
\end{equation}

\begin{figure}
    \centering
    \includegraphics[scale=0.49]{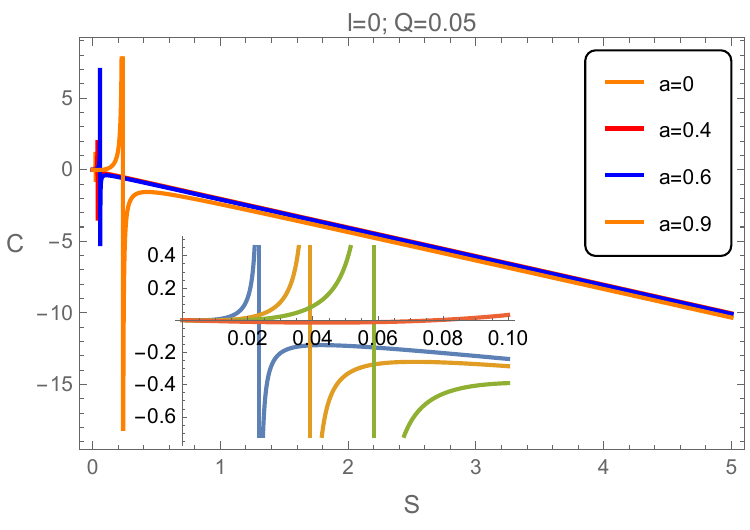}
    \label{im4afrolov} 
    \includegraphics[scale=0.49]{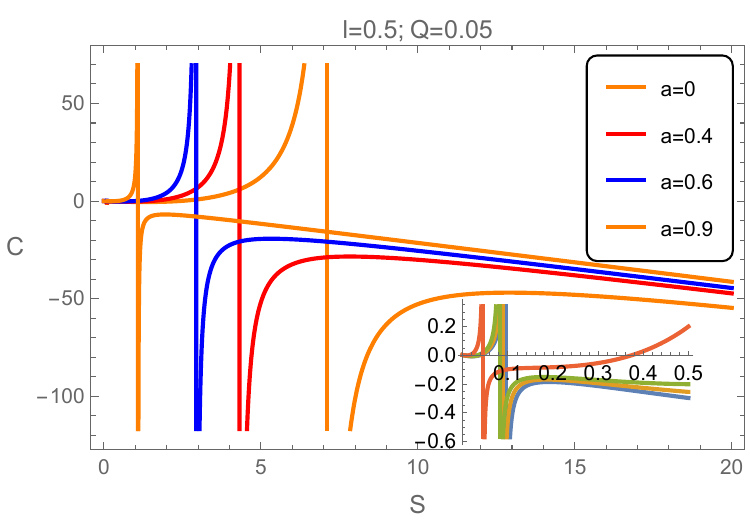}
    \label{im4bfrolov} 
    \includegraphics[scale=0.49]{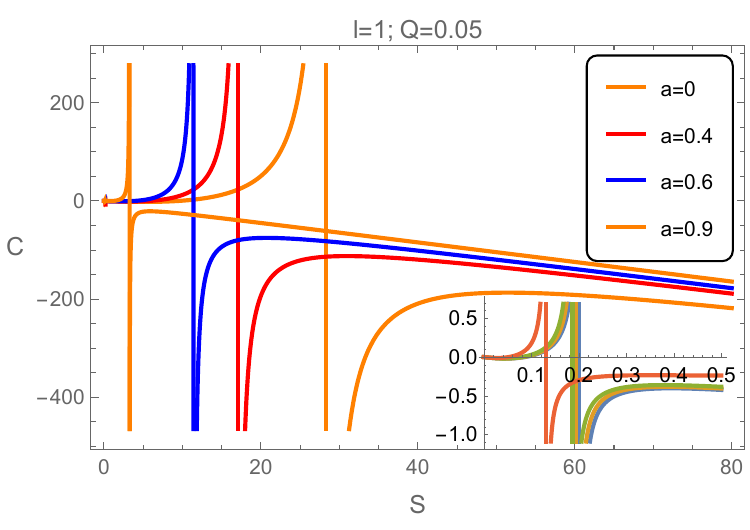}
    \label{im4cfrolov} 
    \includegraphics[scale=0.49]{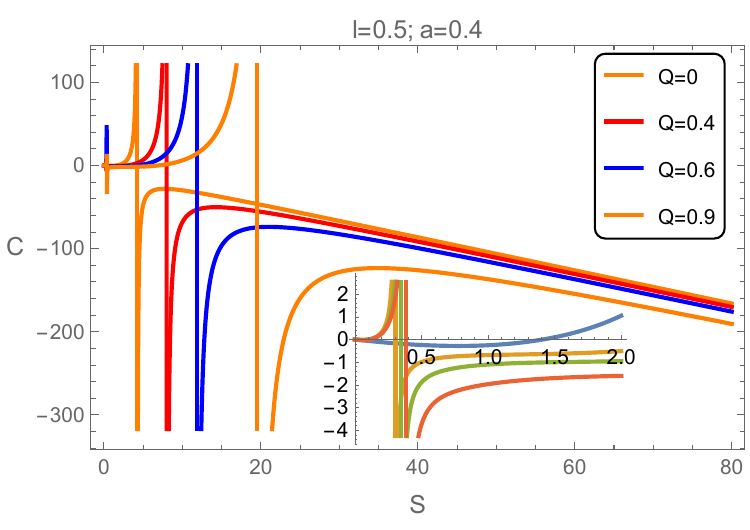}
    \label{im4dfrolov} 
    \caption{Heat capacity as a function of entropy $C(S)$ for different values of $a$, $Q$ and $l$.}
    \label{im4frolovcordas}
  \end{figure}

The behavior of heat capacity as a function of entropy, for different values of $l$, $Q$ and $a$, is given in Fig. \ref{im4frolovcordas}. For $l=0$ and $a=0$, Eq.(\ref{eq:1.70frolovcordas}) reduces to $C=-2S \left(\pi Q^2-S\right)/(3 \pi Q^2-S)$, as expected for the case of Reissner-Nordström spacetime. In this case, there are entropy values for which the heat capacity is positive for $S>0$, indicating a stable thermodynamic system.

When we consider Frolov ($l=1$, $Q=0.05$ and $a=0$) and Frolov spacetime with the cloud of strings ($l=1$, $Q=0.05$ and $0<a<1$), there are values for the entropy for which the heat capacity
acquires positive or negative values. This means that the Frolov black hole can be unstable or stable depending on the entropy values considered. It is important to note that the string parameter plays an important role in the behavior of the heat capacity, modifying the phase transition point. Thus, similar to the Reissner-Nordström black hole, the Frolov black hole and the Frolov black hole with a cloud of strings have regions of stability, depending on the parameters chosen.

\section{Concluding remarks}
\label{sec5}

we generalized the Frolov solution by incorporating a source corresponding to a cloud of strings and examine the  implications associated with the presence of cloud of strings, in the framework of the Frolov black hole with respect to several features of its physics.

Initially, when examining the singularity versus regularity of Frolov spacetime when surounded by a cloud of strings,
using the Kretschmann scalar, we find that the presence of this cloud  breaks the regularity of the original Frolov's black hole solution. 

However, based on the analysis of the geodesic presented in Fig. (\ref{testandofrolovcs}), we observe that, both in Frolov spacetime with a cloud of strings and in the case without a cloud of strings, a test particle can cross the potential barrier and reach the point $r=0$ in a finite time. This result implies the completeness of the geodesics and, consequently, the regularity of spacetime.

Thus, the investigation of geodesics, with the aim of understanding the properties of spacetime near the origin of the black hole, can corroborate or refute the predictions associated with Kretschmann scalar. However, to confirm the presence of singularities in space-time, it is essential to assess geodesic completeness (or its incompleteness), since the evaluation of the Kretschmann scalar offers only a local perspective on the possible singularities of the metric.

With regard to energy conditions, for Frolov spacetime with cloud of string, it is found that energy conditions can be maintained or violated depending on the relationships between black hole parameters. For large radial distances, the analysis indicates that all conditions are satisfied and tend to zero.

We also investigated the thermodynamics of black holes with regard to the behavior of black hole mass as a function of entropy, Hawking temperature, and heat capacity.

With regard to the mass of the black hole, it can be seen in Fig. \ref{im2frolovcordas} that, for the Reissner-Nordström black hole ($a=0$ and $l=0$), the mass parameter only has positive values for positive values of entropy, $S$. Similar behavior is obtained for the Reissner-Nordström black hole ($l=0$) with cloud of strings ($0<a<1$). In the case of the Frolov black hole ($l=0.5$, $Q=0.05$) with ($0<a<1$) and without ($a=0$) cloud of strings, the mass parameter has positive and negative values, depending on the black hole parameters. It should be noted that the cloud of strings parameter modifies the phase transition point for the Frolov black hole surrounded by this cloud.

In Figure \ref{im3frolovcordas}, we show the behavior of temperature as a function of entropy in different scenarios. It can be observed that, in Reissner–Nordström spacetime ($a=0$ and $l=0$), the temperature can assume positive or negative values for $S>0$. Similarly, in the case of Reissner–Nordström–Letelier ($l=0$ and $0<a<1$), the same pattern is observed. Furthermore, when considering Frolov spacetime ($l=1$, $Q=0.05$, and $a=0$), it is noted that $T$ can also be either positive or negative, depending on the black hole parameters. The same behavior is observable when analyzing the cloud of strings in Frolov spacetime ($l=1$, $Q=0.05$, and $0<a<1$).

Finally, we analyze the stability of the black hole by calculating the heat capacity as a function of entropy for different values of $l$, $Q$, and $a$, as shown in Fig. \ref{im4frolovcordas}. In particular, when $l=0$ and $a=0$, Eq.(\ref{eq:1.70frolovcordas}) reduces to $C=-2S \left(\pi Q^2-S\right)/(3 \pi Q^2-S)$, as expected for the case of Reissner-Nordström spacetime. In this situation, there are entropy values with $S>0$ that make the heat capacity positive, signaling a thermodynamically stable system. Considering Frolov spacetime ($l=1$, $Q=0.05$ and $a=0$) and also Frolov spacetime with the cloud of strings ($l=1$, $Q=0.05$ and $0<a<1$), we observe entropy ranges in which the heat capacity can be either positive or negative. This behavior indicates that the Frolov black hole can exhibit regimes of instability or stability, depending on the entropy value analyzed.

It is worth noting that the parameter associated with the string has a significant influence on the evolution of heat capacity, altering the phase transition point. As in the case of Reissner–Nordström, Frolov's black hole, as well as Frolov black hole with a cloud of strings, exhibit regions of stability that depend on the parameters chosen.

\section*{Acknowledgements}
V.B. Bezerra is partially
supported by CNPq-Brazil ( Conselho Nacional de Desenvolvimento Científico e Tecnológico) through Research Project No. 307211/2020-7.

F. F. Nascimento and J. C. Rocha acknowledge Departamento de Fisica, Universidade Federal da Paraiba, for hospitality.

\bibliography{ref}


\end{document}